\documentclass{iau} 
\usepackage{graphicx, commath}
\usepackage{natbib}

\title{Different types of star-planet interactions}
\author{A.~A.~Vidotto}
\affiliation{School of Physics, Trinity College Dublin, The University of Dublin, Dublin 2, Ireland \\
email: {\tt aline.vidotto@tcd.ie} }

\pubyear{2020}
\volume{354}
\jname{Solar and Stellar Magnetic Fields: Origins and Manifestations}
\editors{Kosovichev, Strassmeier \& Jardine, eds.}

\begin{document}

\maketitle

\begin{abstract}
Stars and their exoplanets evolve together. Depending on the physical characteristics of these systems, such as age, orbital distance and activity of the host stars, certain types of star-exoplanet interactions can dominate during given phases of the evolution.  Identifying observable signatures of such interactions can provide additional avenues for characterising exoplanetary systems. Here, I review some recent works on star-planet interactions and discuss their observability at different wavelengths across the electromagnetic spectrum. 
\keywords{stars: magnetic fields, stars: winds, outflows, stars: planetary systems}%% add here a maximum of 10 keywords, to be taken form the file <Keywords.txt>
\end{abstract}

\firstsection % if your document starts with a section,

\section{Introduction}
Stellar magnetic fields drive the space weather of exoplanets. In cool dwarf stars, their magnetic activity are responsible for driving stellar winds and coronal mass ejections, and also for generating high-energy irradiation in the extreme ultraviolet (UV) and X-rays. Hence, understanding the host star magnetism is a key ingredient for characterisation of exoplanetary environments. The magnetic properties of cool stars depends on several ingredients, such as their rotation, age, and internal structure \citep{2014MNRAS.441.2361V}.

\begin{figure}
\includegraphics[width=\textwidth]{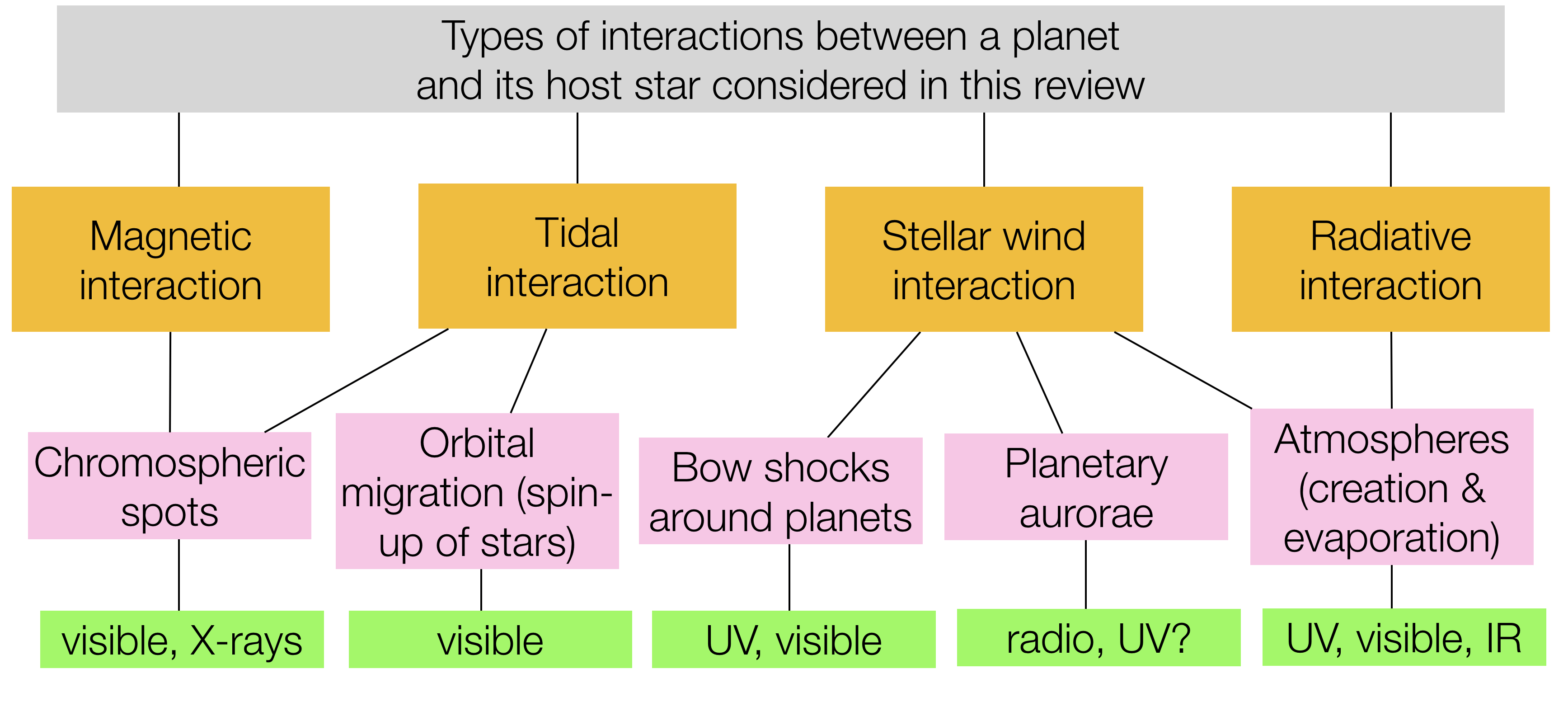}%\label{fig.type}
\end{figure}

Planets that orbit closer to their host stars are embedded in harsher stellar environments, as magnetism and irradiation, for example, decay with some power of the distance. The extreme architecture of most of the known exoplanetary systems, in addition to the differences in magnetic properties of host stars compared to those of our Sun, can give rise to planet-star interactions that are not present in the solar system. These interactions can generate observable signatures, thus providing additional avenues for characterising exoplanetary systems. 

In this review, I present some recent works on star-planet interactions. I consider four different types of interactions, as shown in the diagram (orange boxes): magnetic, tidal, interaction with a stellar wind and with stellar irradiation. Each of these interactions can have different effects on either the star or the planet (pink boxes): causing chromospheric hotspots, migration of orbits, etc. These effects in turn can generate different observables (green boxes) throughout the electromagnetic spectrum: from radio to ultraviolet/X-rays.  

In the Sections that follow, I discuss the effects of different types of star-planet interactions, based on the review talk I presented in the IAU Symposium 354 ``Solar and Stellar Magnetic Fields: Origins and Manifestations''. It is important to note that the signatures of star-planet interactions can occur in the star or in the  planet. These signatures are not necessarily recurrent. In some cases, the signatures are not very strong, which have led to multiple interpretations of the same dataset. The potential material to be reviewed is thus very large! Instead of discussing a particular system in detail, including multiple analyses and interpretations of data, instead, here, I have chosen to present different flavours of interactions between stars and exoplanets. I hope that the reader will appreciate how vibrant this research field is and will use this paper as a starting point to deepen their interest/research in this exciting field!

\section{Anomalous stellar activity}
It did not take long after the discovery of the first exoplanet orbiting a cool dwarf for researchers to propose that close-in planets experimenting strong interactions with their host stars could excite `hot spots' on the chromospheres of their  hosts \citep{2000ApJ...533L.151C}. Spots are seen at the surface of the Sun and indirectly detected in stars, through rotational modulation in their lightcurves. Bright hot spots, on the other hand, are seen in close binary stars, near the sub-binary point. Similarly, the idea is that close-in exoplanets would excited additional, or anomalous, activity in their hosts, in the case of strong star-planet interactions. In both cases though the orbital period and stellar rotational period must differ in order to disentangle the signature of the anomalous activity from that of normal stellar activity. 

Imagine, for example, an exoplanetary system that has tidal interactions similar to the Earth-moon system. As the planet orbits around its host, it raises tidal bulges that are in constant expansion and contraction. The movement of these bulges can dissipate energy and generate heating, that would give rise to anomalous stellar activity. Because two bulges are raised in tidal interactions, this means that  anomalous activity created by tidal interactions should be modulated by half the orbital period of the planet, $P_{\rm orb}/2$ \citep{2001MNRAS.325...55S}. \footnote{In fact, the modulation should occur at the beat period, which is a combination of stellar rotation and orbital periods \citep{2010MNRAS.406..409F}.}

Another example consists of magnetic interactions between a magnetised host star and a magnetised planet. In certain configurations, magnetic field lines of opposite polarities belonging to these two different bodies can  undergo reconnection, releasing energetic particles that travel towards the star along stellar magnetic field lines \citep{2004ApJ...602L..53I}. As these particles impact the stellar chromosphere at the magnetic footpoints, chromospheric hot spots can be excited. As the planet moves through its orbit, the planet interacts with different stellar field lines, so hot spots would be `activated' through the orbit and thus be modulated by the planet's orbital period, $P_{\rm orb}$ \citep{2001MNRAS.325...55S}. \footnote{This scenario implies that the star has a simple magnetic field geometry, while, in reality, the modulation might be more complicated due to complex magnetic field topologies. For example, consider a star with a tilted dipole geometry. This star would have magnetic field lines linking to the planet that are a combination of open lines and closed loops. As the planet moves from one branch of the closed loops to another, the activated chromospheric hot spots would appear to be jumping at different positions in the stellar chromosphere, thus creating a phase lag effect or a jumping effect (e.g., \citealt{2006MNRAS.367L...1M}, \citealt{2019ApJ...881..136S}, Folsom et al., in prep).}

Figure 4 shows that the magnetic field lines linking the planet to the star are a combination of open lines and closed loops. These closed magnetic loops will cause the SPI-related chromospheric spots on the star to move differently from the planetary orbit, with large jumps occurring where the planet moves from one branch of the closed loop to another. This phase lag and jumping effect is evident from Fig. 5 (see also predictions by McIvor et al. 2006). Recently, Strugarek et al. (2019) showed a similar effect in the case of Kepler-78, where the magnetic topol- ogy of the host star can greatly affect the transient nature of SPI. Although their simulations did not explain the amplitude of en- hanced activity observed by Moutou et al. (2016), in stars with stronger magnetic fields (e.g. HD 179949, Fares et al. 2012) the effect may be more detectable.

Hence, the modulation of anomalous activity could tell us something about whether the interaction is of tidal or magnetic type. \citet{2005ApJ...622.1075S, 2008ApJ...676..628S} searched anomalous activity in some close-in planets finding in some cases modulation with $P_{\rm orb}$, and interpreting those as caused by  magnetic interactions, which would allow us to learn about planetary magnetic moment. Recently, \citet{2018AJ....156..262C} investigated anomalous activity in the K dwarf HD189733, triggered by its hot-jupiter. They found that anomalous activity is present in one observational epoch, and is modulated with the orbital period of the planet. This epoch happens to be  when the stellar magnetic field is observed to be stronger \citep{2017MNRAS.471.1246F}, indicating that the level of stellar magnetic activity would either affect when the hotspots are triggered or at the very least when they are detectable. 

An important concept to keep in mind in these interactions is the orbital distance in relation to the Alfven surface of the star. The Alfven surface delineates the regime of the stellar wind that is dominated by magnetic forces (within the surface) or by inertial forces (beyond the surface). Inside the Alfven surface, the stellar wind is sub-Alfvenic, and Alfven ``wing'' currents, connecting star and planet, develop \citep{2015ApJ...815..111S}. However, in the super-Alfvenic regime, this direct connection no longer exists. This implies that, as the stellar magnetism evolves (e.g., through an activity cycle), so does the Alfven surface and the star-planet connectivity \citep{2016MNRAS.459.1907N,2019MNRAS.485.4529K}. This evolution might, for example, affect when anomalous activity through star-planet interaction is triggered, giving rise to an `on/off' nature \citep{2008ApJ...676..628S}. 

This `on/off' regime can also happen in orbital timescales. Imagine the case of an eccentric system: at periastron, the planet interacts with a larger stellar magnetic field, hence there is a stronger star-planet interaction. At apastron, the opposite happens: the large distance means that the stellar magnetic field is weaker, hence would generate a weaker star-planet interaction. The eccentric planet HD17156b was observed at different orbital phases in X-rays by \citet{2015ApJ...811L...2M}, who found enhanced stellar emission a few hours after periastron. This enhanced stellar emission could be interpreted as a magnetic reconnection event that was triggered (or became more powerful) at close orbital separation. Alternatively, it could also be interpreted as some sort of planetary material that evaporated and was accreted into the star, generating a bright X-ray event. Similar to other  works on star-planet interaction, the X-ray  emission was occasional, and not seen in all observed epochs. 

{\it $\bullet $ In the electromagnetic spectrum:  anomalous modulations are usually observed in the visible, with ground-based spectrographs. Observations in X-rays have recently been conducted with  missions like XMM and Chandra and, in the future, with could be possible to conduct with Athena.}

\section{Interactions with stellar winds}
As the wind outflows from the star, it permeates the interplanetary medium. Stellar winds consist of particles and magnetic fields, similar to the solar wind. There are, however, important differences between the solar wind around the Earth and a stellar wind around a close-in exoplanet. Compared to the Earth, the close-in location of hot-Jupiters implies they interact with (1) higher density wind, (2) higher ambient magnetic fields, (3) in general, lower wind velocities, as close-in planets are `parked' inside the acceleration zone of stellar winds. Even though the stellar wind is expected to have lower velocities around close-in planets, the fact that the orbital velocities of these planets are higher imply that, in the reference frame of a close-in exoplanet, the relative velocity between the planet and the host star wind can reach several km/s, going above local sonic speeds (or Alfven speeds) \citep{2011MNRAS.411L..46V}. Overall, the extreme conditions of local stellar winds around close-in exoplanets imply stronger planet-wind interactions than those seen in the solar system \citep{2015MNRAS.449.4117V}. 

In super-sonic (or super-Alfvenic) interactions, ie, bow shocks can develop around the planets \citep{2010ApJ...722L.168V, 2019MNRAS.489.5784C}. In addition, depending on the characteristics of the system (ie, physical conditions of the stellar winds and the planets),  evaporated planetary material may trail the planet \citep{2018MNRAS.479.3115V,2019MNRAS.483.2600D}, forming a comet-like tail structure. Alternatively, in some configurations, material can be funnelled towards the star, forming an inspiralling accretion stream  \citep{2015A&A...578A...6M}.

Observations of these stellar-wind--planet interactions have been done mostly in ultraviolet lines, observed during transits \citep[e.g.][]{2010ApJ...714L.222F, 2016A&A...591A.121B}. In the case of the ultra-hot Jupiter WASP-12b, HST observations demonstrated that the planet starts transiting in the UV much before  the geometric transit \citep{2010ApJ...714L.222F}. This asymmetric transit requires the presence of asymmetric material around the planet. One interpretation is that this asymmetry is created by shocked stellar wind material ahead of the planetary orbit. Contrary to the bow shock surrounding the Earth, which is oriented towards the Sun, the high orbital velocity of WASP-12b causes the bow shock to swing around, appearing at an angle ahead of its orbit. If the bow shock is formed around the magnetosphere of the planet (similar to Earth's bow shock), then these observations can constrain the magnetic field of the planet \citep{2011MNRAS.416L..41L}. 

These observations also have the potential to reveal the physical conditions of stellar winds. In the case of the warm-Neptune GJ436b, models of the asymmetric distribution of material around the planet can constrain local stellar wind conditions \citep{2016A&A...591A.121B}, which can then be used to probe the global properties of stellar winds \citep{2017MNRAS.470.4026V}. Winds of cool dwarfs are very rarefied and often challenging to be observed \citep{2004LRSP....1....2W,2017A&A...599A.127F,2019MNRAS.483..873O,2019MNRAS.482.2853J}. 

In the solar system, the interaction between the solar wind and a planet's magnetosphere give rise not only to a bow shock, but also to radio (auroral) emission. The ``radio Bode's law'' is an empirical finding that shows that the power emitted by the magnetised solar system planet in radio is proportional to the power dissipated by the solar wind on the interaction with the planet \citep{1998JGR...10320159Z}. Extrapolations of this empirical scaling have been used to predict the amount of radio power a close-in exoplanet would emit. The predictions often result in radio powers that are several orders of magnitude larger than that of Jupiter, but so far, most searches for exoplanetary radio emission have come out empty. A hint of detection was reported in \citet{2013A&A...552A..65L} from the hot-Jupiter HAT-P-11b. In this work, it was not the emission of the planet that was seen, but rather the reduced radio emission of the system at the phases where the planet was occulted by the star (secondary transit). Because these are cyclotron emissions, whose frequencies are proportional to the planetary magnetic field, their detections would allow us to derive the magnetic properties of exoplanets. Magnetic fields are windows towards the interior of the planets, as they are generated by dynamo mechanism in planetary interiors. %\citet{2014A&A...562A.108S}

{\it $\bullet $ In the electromagnetic spectrum: UV observations are currently being done with HST. The cubesat CUTE (PI: K.~France;  \citealt{2017SPIE10397E..1AF}) is a dedicated mission  to study spectroscopic transits of close-in exoplanets and will be launched in 2020. Radio observations of planetary radio emission have been conducted with a wide variety of radio telescopes. It is believed that low-frequency observations are more appropriate - if these exoplanets  host 1 -- 100G magnetic field, they would generate emission at cyclotron frequencies of 2.8 -- 280 MHz. LOFAR and, in the future, SKA could be appropriate for detecting these signals.}

%%%%%%%%%%%%%%%%%%%%%%%%%%
\section{Spin-up of stellar rotation}

The orbits of planets can change in time due to several mechanisms, such as tidal interactions, stellar winds that carry away stellar angular momentum (or, conversely, mass accretion that increases stellar angular momentum), friction and planet evaporation. As a consequence, orbital migration could lead a planet to be engulfed by its host star. As the planet moves to inner orbits, conservation of angular momentum of the system implies that a reduction in orbital angular momentum is accompanied by an increase in stellar angular momentum. Therefore, if a planet is engulfed, an accompanied increase in stellar rotation is expected. The timescale for this to happen, if it ever happens in a given planetary system, depends on several characteristics, from stellar properties (including its internal structure, \citealt{2016A&A...591A..45P}) to planetary properties and orbital parameters. 

In studies investigating spin up of stars in the red giant phase, \citet{2016A&A...593L..15P,2016A&A...591A..45P, 2016A&A...593A.128P} suggested that planet engulfment could be the origin of fast rotating red giants. A planet, with mass on the order of 1 to 15 Jupiter masses, initially orbiting at 0.5 au, could spin up the star to equatorial rotation velocities of up to few tens of km/s. These high velocities cannot be explained by models that only take into account the evolution of a single star. In line with what happens in the main sequence \citep{2014MNRAS.441.2361V}, rotation and magnetic fields are linked \citep{2015A&A...574A..90A} are also linked in the post-main sequence. Thus, it is possible that a  high surface magnetic field in red giants could be a detectable signature of planet engulfment \citep{2016A&A...593L..15P}.

Similarly, a close-in planet engulfed during the stellar main-sequence phase could generate high stellar rotation. \citet{2019A&A...621A.124B} investigated potential physical parameters in which an inward migrating planet could spin up a main-sequence solar mass star. They showed that lighter planets ($\lesssim 1M_{\rm jupiter}$), and/or planets with orbits greater than $\sim ~0.03$au, would not spin up their hosts during the main sequence. But ultra hot Jupiters, i.e, more massive planets and at closer orbital distances,  could. 

Stellar rotation also plays a role in planetary migration \citep{2008MNRAS.389.1233L}: a change in sign in tidal forces occur at the corotating radius, which is given by
\begin{equation}
r_{\rm co} = \left( \frac{GM_\star P_{\star, {\rm rot}}^2}{4\pi^2}\right)^{1/3} = 0.02 ~{\rm au} \left(\frac{P_{\star, {\rm rot}}}{1~ \rm day}\right)^{2/3}  \left ( \frac{M_\star}{M_\odot}\right)^{1/3} %4.2 R_\odot
\end{equation}
where $P_{\star, {\rm rot}}$ is the stellar rotation period, $M_\star$ its mass and $G$ is the gravitational constant. A fast rotating star, has a lower corotating radius. For example, at a 1-day period rotation, the corotating radius is at 0.02~au for a solar mass star. Thus, a planet orbiting outside this radius gains angular momentum from the stellar spin, causing the star to spin down and the planet to be pushed away from the star \citep{2008MNRAS.389.1233L,2009ApJ...703.1734V,2010ApJ...720.1262V}. Thus, stars born as fast rotators could push away their inner planets at the beginning of their lives. The farther the planet is, however, means reduced tidal forces, until any orbital evolution becomes insignificant. \citet{2019A&A...621A.124B} showed that as the star spins down during its main-sequence phase, the corotating radius increases and a planet, initially orbiting beyond $r_{\rm co}$, could end up orbiting below $r_{\rm co}$. In this case, even though the transfer of angular momentum would happen from the orbit to the star, the forces might be too low to change the orbital evolution significantly. They showed that, planets orbiting fast rotating hosts would take longer to be engulfed (and could never be engulfed in fact), regardless of its mass. % is expected that higher initial stellar rotation rates lead to longer planet life- times 

{\it $\bullet $ In the electromagnetic spectrum: detecting rotation rates can be done from the ground and from space. We have seen a boost in rotational data with planet detection surveys (Kepler, K2, TESS and in the future Plato, Cheops) and recently with GAIA.  I have the impression that it might be difficult to attribute an `atypical' stellar rotation (i.e., that related to a planet engulfment) in the main sequence, but this could be more easily disentangled in the red giant phase \citep{2016A&A...593L..15P}. Another possible signature of a planet engulfment could be a change in stellar metallicity (``pollution''), possibly detected using ground-based spectrographs. However, I am not aware how much metallicity change to expect nor how long the signature would remain until planetary material becomes mixed with stellar material and, effectively, disappears.  }

%%%%%%%%%%%%%%%%%%%%%%%%
\section{Atmospheric creation or evaporation}

The interaction between a stellar wind and a planetary atmosphere has long been attributed to causing planet erosion. This is, for example, what is believed to have happened to Mars, which does not have an intrinsic magnetic field, therefore lacking a protective `umbrella' against the solar wind. Recently, there has been a debate in the literature whether this is indeed the case, with some authors arguing that magnetic fields do not affect atmospheric escape, with Mars and Earth being counter examples of unmagnetised and magnetised planets, respectively, with similar outflow rates  (\citealt{2010AGUFMSM33B1893S},  see also \citealt{2018MNRAS.481.5146B,2019MNRAS.488.2108E}). Although escape in solar system planets can help us understand exoplanet evaporation (or atmospheric survival), the different, and often very extreme, architectures of exoplanetary systems compared to the solar system does not necessary guarantee that the same evaporation mechanisms taking place in the solar system would operate (or be as strong) in the exoplanets knows to date. 

For example, HST observations have shown that  close-in giant planets  have huge atmospheric escape rates \citep[e.g.][]{2003Natur.422..143V,  2010ApJ...714L.222F, 2015Natur.522..459E, 2016A&A...591A.121B,2014ApJ...786..132K}. \footnote{More recently, escape has also been studied in the infrared triplet line of neutral Helium at 10830\AA\ \citep{2018Sci...362.1388N,2018Natur.557...68S,2019A&A...623A..58A}
 A more indirect detection of planetary mass loss comes from \textit{Kepler} observations of the  distributions of planetary radii and orbital periods (Fig.\,1, \citealt{2013ApJ...763...12B, 2016A&A...589A..75M, 2017AJ....154..109F}), which showed that planetary evaporation is necessary to explain the depletion of small planets with short orbital periods  \citep{2014ApJ...795...65J, 2016MNRAS.455L..96H}. }
  The only mechanism that can generate such a large outflow rate is that of hydrodynamic escape caused by stellar extreme UV (EUV) irradiation. This mechanism is not currently important in the solar system planets.  Irradiation influences exoplanetary atmospheric temperatures and affects mass loss. Due to geometrical effects, close-in planets receive large levels of irradiation from their stars. This increased heating causes their atmospheres to inflate and more likely to outflow through a hydrodynamic escape mechanism. Given that stellar EUV radiation changes through stellar evolution, this means that the survival of atmospheres depends on the EUV history of host star \citep{2015A&A...577L...3T,2015ApJ...815L..12J}. EUV fluxes are related to stellar activity, which is observed to decrease with age. Therefore, a close-in giant planet  orbiting a young star will have a higher outflow rate, which will decrease with the evolution of stellar activity. In comparative terms, a lower gravity giant cannot hold on to its atmosphere in a same way as a higher gravity giant can. The two processes combined thus indicate that young, Saturn-mass planets are more easily to evaporate than older, Jupiter-mass planets orbiting at close distances to their host stars \citep{2019MNRAS.490.3760A}.
 
 In a more counter-intuitive mechanism, \citet{2018MNRAS.481.5296V} showed that planetary atmospheres can be {\it created} (not eroded!) in the interaction with stellar winds. The proposed mechanism is that of sputtering caused by stellar wind particles incident on the bare surface of close-in terrestrial planets. This is modelled in a similar way as to sputtering in Mercury \citep{2015P&SS..115...90P}, but occur with a few orders of magnitude larger incident kinetic energy of the solar wind.  \citet{2018MNRAS.481.5296V} studied atmospheric creation in the close-in terrestrial planets HD219134b and HD219134c, showing that sputtering releases refractory elements from the entire dayside surfaces of the planets, with elements such as O and Mg creating an extended neutral exosphere around these planets. 
 
{\it $\bullet $ In the electromagnetic spectrum: Hydrogen escape has been observed in the UV Ly-$\alpha$ line with HST and some attempts done also in H-$\alpha$, with the benefit that the latter can be conducted from the ground \citep[e.g.][]{2017AJ....153..217C}. Other UV (metal) lines have also tracked escape in observations conducted with HST \citep{2010ApJ...714L.222F} and, in the future, could be done with CUTE \citep{2017SPIE10397E..1AF}.  In the infrared, detections can be made with ground-based spectrographs, such as CARMENES or SpIRou, or from space-based missions, such as HST and, in the future, JWST. }

\section{Conclusions}
In this (too) brief review, I discussed four different types of star-planet interactions: magnetic, tidal, with stellar wind and with stellar radiation. All these interactions can produce observable signatures that can take place in the star or in the planet. Their detections could help further characterise the physical conditions of planetary systems, such as magnetism in exoplanets, stellar wind properties at the orbit of exoplanets, etc. These signatures can take place at different wavelengths across the electromagnetic spectrum, with some new (and old) instruments with great potential for researching star-planet interactions: in the infrared (SpIRou, CARMENES, JWST), low-frequency radio (Lofar, SKA), optical (TESS, Plato, GAIA), UV (HST, CUTE), X-rays (XMM, Chandra, Athena), just to cite a few.  

\section*{Acknowledgements}
I would like to thank (again) the organisers of this fantastic symposium for its organisation, for the invitation to give this review and for partial financial support to attend the meeting. I also acknowledge funding received from the Irish Research Council Laureate Awards 2017/2018, and partially from the European Research Council (ERC) under the European Union's Horizon 2020 research and innovation programme (grant agreement No 817540, ASTROFLOW).

\def\apj{{ApJ}}    
\def\nat{{Nature}}    
\def\jgr{{JGR}}    
\def\apjl{{ApJ Letters}}    
\def\aap{{A\&A}}   
\def\mnras{{MNRAS}}
\def\aj{{AJ}}
\def\apss{{AP\&SS}}
\let\mnrasl=\mnras
\def\planss{{Planetary Space Science}}

%\bibliographystyle{aa}
%\bibliography{../../../Work/artigos/00bibtex-list,other_refs}

\begin{thebibliography}{62}
\expandafter\ifx\csname natexlab\endcsname\relax\def\natexlab#1{#1}\fi

\bibitem[{{Allan} \& {Vidotto}(2019)}]{2019MNRAS.490.3760A}
{Allan}, A. \& {Vidotto}, A.~A. 2019, \mnras, 490, 3760

\bibitem[{{Allart} {et~al.}(2019){Allart}, {Bourrier}, {Lovis}, {Ehrenreich},
  {Aceituno}, {Guijarro}, {Pepe}, {Sing}, {Spake}, \&
  {Wyttenbach}}]{2019A&A...623A..58A}
{Allart}, R., {Bourrier}, V., {Lovis}, C., {et~al.} 2019, \aap, 623, A58

\bibitem[{{Auri{\`e}re} {et~al.}(2015){Auri{\`e}re}, {Konstantinova-Antova},
  {Charbonnel}, {Wade}, {Tsvetkova}, {Petit}, {Dintrans}, {Drake}, {Decressin},
  {Lagarde}, {Donati}, {Roudier}, {Ligni{\`e}res}, {Schr{\"o}der},
  {Landstreet}, {L{\`e}bre}, {Weiss}, \& {Zahn}}]{2015A&A...574A..90A}
{Auri{\`e}re}, M., {Konstantinova-Antova}, R., {Charbonnel}, C., {et~al.} 2015,
  \aap, 574, A90

\bibitem[{{Beaug{\'e}} \& {Nesvorn{\'y}}(2013)}]{2013ApJ...763...12B}
{Beaug{\'e}}, C. \& {Nesvorn{\'y}}, D. 2013, \apj, 763, 12

\bibitem[{{Benbakoura} {et~al.}(2019){Benbakoura}, {R{\'e}ville}, {Brun}, {Le
  Poncin-Lafitte}, \& {Mathis}}]{2019A&A...621A.124B}
{Benbakoura}, M., {R{\'e}ville}, V., {Brun}, A.~S., {Le Poncin-Lafitte}, C., \&
  {Mathis}, S. 2019, \aap, 621, A124

\bibitem[{{Blackman} \& {Tarduno}(2018)}]{2018MNRAS.481.5146B}
{Blackman}, E.~G. \& {Tarduno}, J.~A. 2018, \mnras, 481, 5146

\bibitem[{{Bourrier} {et~al.}(2016){Bourrier}, {Lecavelier des Etangs},
  {Ehrenreich}, {Tanaka}, \& {Vidotto}}]{2016A&A...591A.121B}
{Bourrier}, V., {Lecavelier des Etangs}, A., {Ehrenreich}, D., {Tanaka}, Y.~A.,
  \& {Vidotto}, A.~A. 2016, \aap, 591, A121

\bibitem[{{Carolan} {et~al.}(2019){Carolan}, {Vidotto}, {Loesch}, \&
  {Coogan}}]{2019MNRAS.489.5784C}
{Carolan}, S., {Vidotto}, A.~A., {Loesch}, C., \& {Coogan}, P. 2019, \mnras,
  489, 5784

\bibitem[{{Cauley} {et~al.}(2017){Cauley}, {Redfield}, \&
  {Jensen}}]{2017AJ....153..217C}
{Cauley}, P.~W., {Redfield}, S., \& {Jensen}, A.~G. 2017, \aj, 153, 217

\bibitem[{{Cauley} {et~al.}(2018){Cauley}, {Shkolnik}, {Llama}, {Bourrier}, \&
  {Moutou}}]{2018AJ....156..262C}
{Cauley}, P.~W., {Shkolnik}, E.~L., {Llama}, J., {Bourrier}, V., \& {Moutou},
  C. 2018, \aj, 156, 262

\bibitem[{{Cuntz} {et~al.}(2000){Cuntz}, {Saar}, \&
  {Musielak}}]{2000ApJ...533L.151C}
{Cuntz}, M., {Saar}, S.~H., \& {Musielak}, Z.~E. 2000, \apjl, 533, L151

\bibitem[{{Daley-Yates} \& {Stevens}(2019)}]{2019MNRAS.483.2600D}
{Daley-Yates}, S. \& {Stevens}, I.~R. 2019, \mnras, 483, 2600

\bibitem[{{Egan} {et~al.}(2019){Egan}, {Jarvinen}, {Ma}, \&
  {Brain}}]{2019MNRAS.488.2108E}
{Egan}, H., {Jarvinen}, R., {Ma}, Y., \& {Brain}, D. 2019, \mnras, 488, 2108

\bibitem[{{Ehrenreich} {et~al.}(2015){Ehrenreich}, {Bourrier}, {Wheatley}, {Des
  Etangs}, {H{\'e}brard}, {Udry}, {Bonfils}, {Delfosse}, {D{\'e}sert}, {Sing},
  \& {Vidal-Madjar}}]{2015Natur.522..459E}
{Ehrenreich}, D., {Bourrier}, V., {Wheatley}, P.~J., {et~al.} 2015, \nat, 522,
  459

\bibitem[{{Fares} {et~al.}(2017){Fares}, {Bourrier}, {Vidotto}, {Moutou},
  {Jardine}, {Zarka}, {Helling}, {Lecavelier des Etangs}, {Llama}, {Louden},
  {Wheatley}, \& {Ehrenreich}}]{2017MNRAS.471.1246F}
{Fares}, R., {Bourrier}, V., {Vidotto}, A.~A., {et~al.} 2017, \mnras, 471, 1246

\bibitem[{{Fares} {et~al.}(2010){Fares}, {Donati}, {Moutou}, {Jardine},
  {Grie{\ss}meier}, {Zarka}, {Shkolnik}, {Bohlender}, {Catala}, \&
  {Cameron}}]{2010MNRAS.406..409F}
{Fares}, R., {Donati}, J., {Moutou}, C., {et~al.} 2010, \mnras, 406, 409

\bibitem[{{Fichtinger} {et~al.}(2017){Fichtinger}, {Guedel}, {Mutel},
  {Hallinan}, {Gaidos}, {Skinner}, {Lynch}, \& {Gayley}}]{2017A&A...599A.127F}
{Fichtinger}, B., {Guedel}, M., {Mutel}, R.~L., {et~al.} 2017, \aap, 599, A127

\bibitem[{{Fleming} {et~al.}(2017){Fleming}, {France}, {Nell}, {Kohnert},
  {Pool}, {Egan}, {Fossati}, {Koskinen}, {Vidotto}, {Hoadley}, {Desert},
  {Beasley}, \& {Petit}}]{2017SPIE10397E..1AF}
{Fleming}, B.~T., {France}, K., {Nell}, N., {et~al.} 2017, in Society of
  Photo-Optical Instrumentation Engineers (SPIE) Conference Series, Vol. 10397,
  Society of Photo-Optical Instrumentation Engineers (SPIE) Conference Series,
  103971A

\bibitem[{{Fossati} {et~al.}(2010){Fossati}, {Haswell}, {Froning}, {Hebb},
  {Holmes}, {Kolb}, {Helling}, {Carter}, {Wheatley}, {Cameron}, {Loeillet},
  {Pollacco}, {Street}, {Stempels}, {Simpson}, {Udry}, {Joshi}, {West},
  {Skillen}, \& {Wilson}}]{2010ApJ...714L.222F}
{Fossati}, L., {Haswell}, C.~A., {Froning}, C.~S., {et~al.} 2010, \apjl, 714,
  L222

\bibitem[{{Fulton} {et~al.}(2017){Fulton}, {Petigura}, {Howard}, {Isaacson},
  {Marcy}, {Cargile}, {Hebb}, {Weiss}, {Johnson}, {Morton}, {Sinukoff},
  {Crossfield}, \& {Hirsch}}]{2017AJ....154..109F}
{Fulton}, B.~J., {Petigura}, E.~A., {Howard}, A.~W., {et~al.} 2017, \aj, 154,
  109

\bibitem[{{Helled} {et~al.}(2016){Helled}, {Lozovsky}, \&
  {Zucker}}]{2016MNRAS.455L..96H}
{Helled}, R., {Lozovsky}, M., \& {Zucker}, S. 2016, \mnras, 455, L96

\bibitem[{{Ip} {et~al.}(2004){Ip}, {Kopp}, \& {Hu}}]{2004ApJ...602L..53I}
{Ip}, W.-H., {Kopp}, A., \& {Hu}, J.-H. 2004, \apjl, 602, L53

\bibitem[{{Jardine} \& {Collier Cameron}(2019)}]{2019MNRAS.482.2853J}
{Jardine}, M. \& {Collier Cameron}, A. 2019, \mnras, 482, 2853

\bibitem[{{Jin} {et~al.}(2014){Jin}, {Mordasini}, {Parmentier}, {van Boekel},
  {Henning}, \& {Ji}}]{2014ApJ...795...65J}
{Jin}, S., {Mordasini}, C., {Parmentier}, V., {et~al.} 2014, \apj, 795, 65

\bibitem[{{Johnstone} {et~al.}(2015){Johnstone}, {Guedel}, {St{\"o}kl},
  {Lammer}, {Tu}, {Kislyakova}, {L{\"u}ftinger}, {Odert}, {Erkaev}, \&
  {Dorfi}}]{2015ApJ...815L..12J}
{Johnstone}, C.~P., {Guedel}, M., {St{\"o}kl}, A., {et~al.} 2015, \apjl, 815,
  L12

\bibitem[{{Kavanagh} {et~al.}(2019){Kavanagh}, {Vidotto},
  {{\'O}.~Fionnag{\'a}in}, {Bourrier}, {Fares}, {Jardine}, {Helling}, {Moutou},
  {Llama}, \& {Wheatley}}]{2019MNRAS.485.4529K}
{Kavanagh}, R.~D., {Vidotto}, A.~A., {{\'O}.~Fionnag{\'a}in}, D., {et~al.}
  2019, \mnras, 485, 4529

\bibitem[{{Kulow} {et~al.}(2014){Kulow}, {France}, {Linsky}, \&
  {Loyd}}]{2014ApJ...786..132K}
{Kulow}, J.~R., {France}, K., {Linsky}, J., \& {Loyd}, R.~O.~P. 2014, \apj,
  786, 132

\bibitem[{{Lecavelier des Etangs} {et~al.}(2012){Lecavelier des Etangs},
  {Bourrier}, {Wheatley}, {Dupuy}, {Ehrenreich}, {Vidal-Madjar}, {H{\'e}brard},
  {Ballester}, {D{\'e}sert}, {Ferlet}, \& {Sing}}]{2012A&A...543L...4L}
{Lecavelier des Etangs}, A., {Bourrier}, V., {Wheatley}, P.~J., {et~al.} 2012,
  \aap, 543, L4

\bibitem[{{Lecavelier des Etangs} {et~al.}(2013){Lecavelier des Etangs},
  {Sirothia}, {Gopal-Krishna}, \& {Zarka}}]{2013A&A...552A..65L}
{Lecavelier des Etangs}, A., {Sirothia}, S.~K., {Gopal-Krishna}, \& {Zarka}, P.
  2013, \aap, 552, A65

\bibitem[{{Llama} {et~al.}(2011){Llama}, {Wood}, {Jardine}, {Vidotto},
  {Helling}, {Fossati}, \& {Haswell}}]{2011MNRAS.416L..41L}
{Llama}, J., {Wood}, K., {Jardine}, M., {et~al.} 2011, \mnrasl, 416, L41

\bibitem[{{Lovelace} {et~al.}(2008){Lovelace}, {Romanova}, \&
  {Barnard}}]{2008MNRAS.389.1233L}
{Lovelace}, R.~V.~E., {Romanova}, M.~M., \& {Barnard}, A.~W. 2008, \mnras, 389,
  1233

\bibitem[{{Maggio} {et~al.}(2015){Maggio}, {Pillitteri}, {Scandariato},
  {Lanza}, {Sciortino}, {Borsa}, {Bonomo}, {Claudi}, {Covino}, {Desidera},
  {Gratton}, {Micela}, {Pagano}, {Piotto}, {Sozzetti}, {Cosentino}, \&
  {Maldonado}}]{2015ApJ...811L...2M}
{Maggio}, A., {Pillitteri}, I., {Scandariato}, G., {et~al.} 2015, \apjl, 811,
  L2

\bibitem[{{Matsakos} {et~al.}(2015){Matsakos}, {Uribe}, \&
  {K{\"o}nigl}}]{2015A&A...578A...6M}
{Matsakos}, T., {Uribe}, A., \& {K{\"o}nigl}, A. 2015, \aap, 578, A6

\bibitem[{{Mazeh} {et~al.}(2016){Mazeh}, {Holczer}, \&
  {Faigler}}]{2016A&A...589A..75M}
{Mazeh}, T., {Holczer}, T., \& {Faigler}, S. 2016, \aap, 589, A75

\bibitem[{{McIvor} {et~al.}(2006){McIvor}, {Jardine}, \&
  {Holzwarth}}]{2006MNRAS.367L...1M}
{McIvor}, T., {Jardine}, M., \& {Holzwarth}, V. 2006, \mnrasl, 367, L1

\bibitem[{{Nicholson} {et~al.}(2016){Nicholson}, {Vidotto}, {Mengel},
  {Brookshaw}, {Carter}, {Petit}, {Marsden}, {Jeffers}, {Fares}, \& {the BCool
  Collaboration}}]{2016MNRAS.459.1907N}
{Nicholson}, B.~A., {Vidotto}, A.~A., {Mengel}, M., {et~al.} 2016, \mnras, 459,
  1907

\bibitem[{{Nortmann} {et~al.}(2018){Nortmann}, {Pall{\'e}}, {Salz},
  {Sanz-Forcada}, {Nagel}, {Alonso-Floriano}, {Czesla}, {Yan}, {Chen},
  {Snellen}, {Zechmeister}, {Schmitt}, {L{\'o}pez-Puertas}, {Casasayas-Barris},
  {Bauer}, {Amado}, {Caballero}, {Dreizler}, {Henning}, {Lamp{\'o}n}, {Montes},
  {Molaverdikhani}, {Quirrenbach}, {Reiners}, {Ribas}, {S{\'a}nchez-L{\'o}pez},
  {Schneider}, \& {Zapatero Osorio}}]{2018Sci...362.1388N}
{Nortmann}, L., {Pall{\'e}}, E., {Salz}, M., {et~al.} 2018, Science, 362, 1388

\bibitem[{{{\'O} Fionnag{\'a}in} {et~al.}(2019){{\'O} Fionnag{\'a}in},
  {Vidotto}, {Petit}, {Folsom}, {Jeffers}, {Marsden}, {Morin}, {do Nascimento},
  \& {BCool Collaboration}}]{2019MNRAS.483..873O}
{{\'O} Fionnag{\'a}in}, D., {Vidotto}, A.~A., {Petit}, P., {et~al.} 2019,
  \mnras, 483, 873

\bibitem[{{Pfleger} {et~al.}(2015){Pfleger}, {Lichtenegger}, {Wurz}, {Lammer},
  {Kallio}, {Alho}, {Mura}, {McKenna-Lawlor}, \&
  {Mart{\'{\i}}n-Fern{\'a}ndez}}]{2015P&SS..115...90P}
{Pfleger}, M., {Lichtenegger}, H.~I.~M., {Wurz}, P., {et~al.} 2015, \planss,
  115, 90

\bibitem[{{Privitera} {et~al.}(2016{\natexlab{a}}){Privitera}, {Meynet},
  {Eggenberger}, {Georgy}, {Ekstr{\"o}m}, {Vidotto}, {Bianda}, {Villaver}, \&
  {ud-Doula}}]{2016A&A...593L..15P}
{Privitera}, G., {Meynet}, G., {Eggenberger}, P., {et~al.} 2016{\natexlab{a}},
  \aap, 593, L15

\bibitem[{{Privitera} {et~al.}(2016{\natexlab{b}}){Privitera}, {Meynet},
  {Eggenberger}, {Vidotto}, {Villaver}, \& {Bianda}}]{2016A&A...591A..45P}
{Privitera}, G., {Meynet}, G., {Eggenberger}, P., {et~al.} 2016{\natexlab{b}},
  \aap, 591, A45

\bibitem[{{Privitera} {et~al.}(2016{\natexlab{c}}){Privitera}, {Meynet},
  {Eggenberger}, {Vidotto}, {Villaver}, \& {Bianda}}]{2016A&A...593A.128P}
{Privitera}, G., {Meynet}, G., {Eggenberger}, P., {et~al.} 2016{\natexlab{c}},
  \aap, 593, A128

\bibitem[{{Saar} \& {Cuntz}(2001)}]{2001MNRAS.325...55S}
{Saar}, S.~H. \& {Cuntz}, M. 2001, \mnras, 325, 55

\bibitem[{{Shkolnik} {et~al.}(2008){Shkolnik}, {Bohlender}, {Walker}, \&
  {Collier Cameron}}]{2008ApJ...676..628S}
{Shkolnik}, E., {Bohlender}, D.~A., {Walker}, G.~A.~H., \& {Collier Cameron},
  A. 2008, \apj, 676, 628

\bibitem[{{Shkolnik} {et~al.}(2005){Shkolnik}, {Walker}, {Bohlender}, {Gu}, \&
  {K{\"u}rster}}]{2005ApJ...622.1075S}
{Shkolnik}, E., {Walker}, G.~A.~H., {Bohlender}, D.~A., {Gu}, P.-G., \&
  {K{\"u}rster}, M. 2005, \apj, 622, 1075

\bibitem[{{Spake} {et~al.}(2018){Spake}, {Sing}, {Evans}, {Oklop{\v{c}}i{\'c}},
  {Bourrier}, {Kreidberg}, {Rackham}, {Irwin}, {Ehrenreich}, {Wyttenbach},
  {Wakeford}, {Zhou}, {Chubb}, {Nikolov}, {Goyal}, {Henry}, {Williamson},
  {Blumenthal}, {Anderson}, {Hellier}, {Charbonneau}, {Udry}, \&
  {Madhusudhan}}]{2018Natur.557...68S}
{Spake}, J.~J., {Sing}, D.~K., {Evans}, T.~M., {et~al.} 2018, \nat, 557, 68

\bibitem[{{Strangeway} {et~al.}(2010){Strangeway}, {Russell}, {Luhmann},
  {Moore}, {Foster}, {Barabash}, \& {Nilsson}}]{2010AGUFMSM33B1893S}
{Strangeway}, R.~J., {Russell}, C.~T., {Luhmann}, J.~G., {et~al.} 2010, in AGU
  Fall Meeting Abstracts, Vol. 2010, SM33B--1893

\bibitem[{{Strugarek} {et~al.}(2019){Strugarek}, {Brun}, {Donati}, {Moutou}, \&
  {R{\'e}ville}}]{2019ApJ...881..136S}
{Strugarek}, A., {Brun}, A.~S., {Donati}, J.~F., {Moutou}, C., \&
  {R{\'e}ville}, V. 2019, \apj, 881, 136

\bibitem[{{Strugarek} {et~al.}(2015){Strugarek}, {Brun}, {Matt}, \&
  {R{\'e}ville}}]{2015ApJ...815..111S}
{Strugarek}, A., {Brun}, A.~S., {Matt}, S.~P., \& {R{\'e}ville}, V. 2015, \apj,
  815, 111

\bibitem[{{Tu} {et~al.}(2015){Tu}, {Johnstone}, {Guedel}, \&
  {Lammer}}]{2015A&A...577L...3T}
{Tu}, L., {Johnstone}, C.~P., {Guedel}, M., \& {Lammer}, H. 2015, \aap, 577, L3

\bibitem[{{Vidal-Madjar} {et~al.}(2003){Vidal-Madjar}, {Lecavelier des Etangs},
  {D{\'e}sert}, {Ballester}, {Ferlet}, {H{\'e}brard}, \&
  {Mayor}}]{2003Natur.422..143V}
{Vidal-Madjar}, A., {Lecavelier des Etangs}, A., {D{\'e}sert}, J.-M., {et~al.}
  2003, \nat, 422, 143

\bibitem[{{Vidotto} \& {Bourrier}(2017)}]{2017MNRAS.470.4026V}
{Vidotto}, A.~A. \& {Bourrier}, V. 2017, \mnras, 470, 4026

\bibitem[{{Vidotto} {et~al.}(2015){Vidotto}, {Fares}, {Jardine}, {Moutou}, \&
  {Donati}}]{2015MNRAS.449.4117V}
{Vidotto}, A.~A., {Fares}, R., {Jardine}, M., {Moutou}, C., \& {Donati}, J.-F.
  2015, \mnras, 449, 4117

\bibitem[{{Vidotto} {et~al.}(2014){Vidotto}, {Gregory}, {Jardine}, {Donati},
  {Petit}, {Morin}, {Folsom}, {Bouvier}, {Cameron}, {Hussain}, {Marsden},
  {Waite}, {Fares}, {Jeffers}, \& {do Nascimento}}]{2014MNRAS.441.2361V}
{Vidotto}, A.~A., {Gregory}, S.~G., {Jardine}, M., {et~al.} 2014, \mnras, 441,
  2361

\bibitem[{{Vidotto} {et~al.}(2010{\natexlab{a}}){Vidotto}, {Jardine}, \&
  {Helling}}]{2010ApJ...722L.168V}
{Vidotto}, A.~A., {Jardine}, M., \& {Helling}, C. 2010{\natexlab{a}}, \apjl,
  722, L168

\bibitem[{{Vidotto} {et~al.}(2011){Vidotto}, {Jardine}, \&
  {Helling}}]{2011MNRAS.411L..46V}
{Vidotto}, A.~A., {Jardine}, M., \& {Helling}, C. 2011, \mnrasl, 411, L46

\bibitem[{{Vidotto} {et~al.}(2018){Vidotto}, {Lichtenegger}, {Fossati},
  {Folsom}, {Wood}, {Murthy}, {Petit}, {Sreejith}, \&
  {Valyavin}}]{2018MNRAS.481.5296V}
{Vidotto}, A.~A., {Lichtenegger}, H., {Fossati}, L., {et~al.} 2018, \mnras,
  481, 5296

\bibitem[{{Vidotto} {et~al.}(2009){Vidotto}, {Opher}, {Jatenco-Pereira}, \&
  {Gombosi}}]{2009ApJ...703.1734V}
{Vidotto}, A.~A., {Opher}, M., {Jatenco-Pereira}, V., \& {Gombosi}, T.~I. 2009,
  \apj, 703, 1734

\bibitem[{{Vidotto} {et~al.}(2010{\natexlab{b}}){Vidotto}, {Opher},
  {Jatenco-Pereira}, \& {Gombosi}}]{2010ApJ...720.1262V}
{Vidotto}, A.~A., {Opher}, M., {Jatenco-Pereira}, V., \& {Gombosi}, T.~I.
  2010{\natexlab{b}}, \apj, 720, 1262

\bibitem[{{Villarreal D'Angelo} {et~al.}(2018){Villarreal D'Angelo},
  {Esquivel}, {Schneiter}, \& {Sgr{\'o}}}]{2018MNRAS.479.3115V}
{Villarreal D'Angelo}, C., {Esquivel}, A., {Schneiter}, M., \& {Sgr{\'o}},
  M.~A. 2018, \mnras, 479, 3115

\bibitem[{{Wood}(2004)}]{2004LRSP....1....2W}
{Wood}, B.~E. 2004, Living Reviews in Solar Physics, 1, 2

\bibitem[{{Zarka}(1998)}]{1998JGR...10320159Z}
{Zarka}, P. 1998, \jgr, 103, 20159

\end{thebibliography}

\begin{discussion}
\discuss{Chia-Hsien Lin}{Could you explain the relation between the increase in planetary evaporation on HD189733b and the presence of a flare/CME?}
\discuss{Vidotto}{The evaporation of the atmosphere of a close-in planet is affected by both the amount of stellar irradiation (which heats the atmosphere of the planet) and the stellar wind/CME (which interacts with the upper atmosphere of a planet). In transit observations of HD189733b performed at the Ly-$\alpha$ line, \citet{2012A&A...543L...4L} noticed an increase in planetary evaporation that occurred about 8 hours after a flare of the host star. Although we do not know if the two events are related, the timing is very suggestive that the flare, somehow, led to an enhance of planet evaporation. There are two possible explanations on how this enhancement took place. For example, it could be that the flare caused an increase in irradiation arriving at the planet atmosphere, which caused an enhance in the evaporation. Alternatively, it could be that, associated to this, there was an ejection of a CME, which interacted with the atmosphere and could have led to an enhance in atmospheric escape.}

\discuss{Kutluay Y\"uce}{What are the most-common spectral types of known planet hosting stars? What do you think about earlier type stars (B,A,F) as potential planet-hosts?}
\discuss{Vidotto}{Due to biases in planet detection techniques, most of the planet hosting stars known today are cool dwarfs with masses $\lesssim 1.3 M_\odot$ and, in most of the cases, these stars tend to be inactive. It is more challenging to detect planets around earlier type stars (A and B spectral types, for instance). This happens because the transit signature is proportional to the ratio $(R_{\rm planet}/R_\star)^2$ (i.e., the area of the planet divided by the area of the star), so the bigger the star, the smaller the signature. In the radial velocity technique, the orbit of the star about the common centre of mass (the reflex motion) is very small and the Doppler shifts caused by the planet motion are thus harder to detect. Having said that, there are now some (a few tens) detections of planets orbiting around earlier type stars, like KELT-9b and WASP-33b, whose hosts are A0 and A5 stars, respectively.}

\discuss{Zhanwen Han}{What are the physical causes of the planet engulfment? Is it due to the expansion of the star in the post-main sequence or the interaction with the stellar wind that causes the loss of orbital angular momentum?}
\discuss{Vidotto}{The reasons why orbital angular momentum changes over time are due to several physical processes, such as tidal interactions, stellar winds that carry away stellar angular momentum (or, conversely, mass accretion that increases stellar angular momentum), friction and planet evaporation. Changes in orbits have an impact on the rotation of the star, which also modifies the structure of the star. This, in turn, also affect the mechanisms of orbital decay. }

\end{discussion}

\end{document}